\documentclass[twocolumn]{aastex61}
\usepackage{enumitem}
\DeclareMathAlphabet{\mathitbf}{OML}{cmm}{b}{it}
\newcommand{\Avec}{\mathitbf{A}}

\newcommand{\Apvec}{\mathitbf{A}_{\mathrm{0}}}
\newcommand{\Bvec}{\mathitbf{B}}
\newcommand{\Bx}{\mathit{B_x}}
\newcommand{\By}{\mathit{B_y}}
\newcommand{\Bz}{\mathit{B_z}}
\newcommand{\Br}{\mathit{B_r}}
\newcommand{\Btheta}{\mathit{B_\theta}}
\newcommand{\Bphi}{\mathit{B_\phi}}
\newcommand{\Bpvec}{\mathitbf{B}_{\mathrm{0}}}

\newcommand{\Etot}{E}
\newcommand{\Ej}{E_{\mathrm{J}}}
\newcommand{\Epot}{E_{\mathrm{0}}}
\newcommand{\Efree}{E_{\mathrm{F}}}
\newcommand{\Ediv}{E_{\mathrm{div}}}
\newcommand{\Ejs}{E_{\mathrm{J},s}}
\newcommand{\Eps}{E_{\mathrm{0},s}}
\newcommand{\Ejns}{E_{\mathrm{J},ns}}
\newcommand{\Epns}{E_{\mathrm{0},ns}}
\newcommand{\Emix}{E_{\mathrm{mix}}}
\newcommand{\hpj}{H_{\mathrm{PJ}}}
\newcommand{\hj}{H_{\mathrm{J}}}
\newcommand{\hv}{H_{\mathcal{V}}}
\newcommand{\efprime}{\bar\Efree/\Etot}
\newcommand{\hjprime}{|\bar\hj|/|\bar\hv|}

\newcommand{\hvprime}{\bar\hv/\phi^2}
\newcommand{\thetaj}{\theta_J}
\newcommand{\fiavg}{\langle|f_i|\rangle}
\newcommand{\dc}{^\circ}
\newcommand{\erg}{~\mathrm{erg}}
\newcommand{\mx}{~\mathrm{Mx}}
\newcommand{\mxmx}{~\mathrm{Mx}^2}
\newcommand{\FVjt}{$\mathrm{FV}_{\rm Coul\_JT}$}
\newcommand{\FVkm}{$\mathrm{FV}_{\rm DeV\_KM}$}
\newcommand{\FVgv}{$\mathrm{FV}_{\rm DeV\_GV}$}
\newcommand{\FVkmw}{$\mathrm{FV}_{\rm DeV\_KM}^{w}$}

\newcommand{\ie}{{\it i.e.}}
\newcommand{\eg}{{\it e.g.}}

\newcommand{\sdo}{{\it SDO}}

\newcommand{\asecs}{\mbox{\ensuremath{^{\prime\prime}}}}
\shortauthors{Thalmann et al.}
\accepted{\today}
\submitjournal{ApJ on Aug 26, 2019}

\begin{document}

\title{Magnetic helicity budget of solar active regions \\ prolific of eruptive and confined flares}

\correspondingauthor{Julia K. Thalmann}
\email{julia.thalmann@uni-graz.at}

\author{Julia K. Thalmann}
\affil{University of Graz, Institute of Physics/IGAM, Universit\"atsplatz 5, 8010 Graz, Austria}

\author{K.~Moraitis}
\author{L.~Linan}
\author{E.~Pariat}
\affiliation{LESIA, Observatoire de Paris, Universit{\'e} PSL, CNRS, Sorbonne Universit{\'e}, Universit{\'e} de Paris, 5 place Jules Janssen, 92195 Meudon, France}

\author{G.~Valori}
\affiliation{Mullard Space Science Laboratory, University College London, Holmbury St.\ Mary, Dorking, Surrey RH5 6NT, UK}

\author{K.~Dalmasse}
\affiliation{IRAP, Universit{\'e} de Toulouse, CNRS, CNES, UPS, 31028 Toulouse, France}

\begin{abstract}
We compare the coronal magnetic energy and helicity of two solar active regions (ARs), prolific in major eruptive (AR~11158) and confined (AR~12192) flaring, and analyze the potential of deduced proxies to forecast upcoming flares. Based on nonlinear force-free (NLFF) coronal magnetic field models with a high degree of solenoidality, and applying three different computational methods to investigate the coronal magnetic helicity, we are able to draw conclusions with a high level of confidence. Based on real observations of two solar ARs we checked trends regarding the potential eruptivity of the active-region corona, as suggested earlier in works that were based on numerical simulations, or solar observations. 
Our results support that the ratio of current-carrying to total helicity, $|\hj|/|\hv|$, shows a strong ability to indicate the eruptive potential of a solar AR. However, $|\hj|/|\hv|$ seems not to be indicative for the magnitude or type of an upcoming flare (confined or eruptive). Interpreted in context with earlier observational studies, our findings furthermore support that the total relative helicity normalized to the magnetic flux at the NLFF model's lower boundary, $\hv/\phi^2$, represents no indicator for the eruptivity.
\end{abstract}


\section{Introduction}\label{sec:introduction} 

Magnetic helicity is uniquely related to the geometrical complexity of the underlying magnetic system, determined by the twist and writhe of individual magnetic field lines, as well as their mutual entanglement. Magnetic helictity is a signed scalar quantity that is (almost) conserved in (resistive) ideal MHD \citep[][]{1984GApFD..30...79B,2015A&A...580A.128P}. Its time evolution reflects the dynamic evolution of the respective magnetic system. For practical cases, such as the solar corona, a gauge-invariant form of the magnetic helicity has been introduced to allow a physically meaningful estimation \citep{1984JFM...147..133B,1984CPPCF...9..111F}, in the form
\begin{equation}
	\hv=\int_\mathcal{V}\left(\Avec+\Apvec\right)\cdot\left(\Bvec-\Bpvec\right)\,{\rm d}\mathcal{V}, \label{eq:hv}
\end{equation}
where the reference field $\Bpvec$ shares the normal component of the studied field $\Bvec$ on the volume's boundary, $\partial\mathcal{V}$. Usually a potential (current-free) field is used as reference field. Here, $\Avec$ and $\Apvec$ are the vector potentials of $\Bvec$ and $\Bpvec$, respectively, where $\Bvec=\nabla\times\Avec$ and $\Bpvec=\nabla\times\Apvec$. 

Since $\hv$ in \href{eq:hv}{Eq.~(\ref{eq:hv})} is computed with respect to a reference field it is called ``relative helicity''.  \cite{2012SoPh..278..347V} demonstrated the validity and physical meaningfulness to compute (and track in time) the relative magnetic helicity in finite volumes in order to characterize (the evolution of) a magnetic system.

Following \cite{1999PPCF...41B.167B}, \href{eq:hv}{Eq.~(\ref{eq:hv})} may be written as $\hv=\hj+\hpj$, with
\begin{eqnarray}
	\hj &=& \int_\mathcal{V}\left(\Avec-\Apvec\right)\cdot\left(\Bvec-\Bpvec\right)\,{\rm d}\mathcal{V},\\
	\hpj &=& \mathrm{~2}\int_\mathcal{V}\Apvec\cdot\left(\Bvec-\Bpvec\right)\,{\rm d}\mathcal{V},
\end{eqnarray}
where, $\hj$ is the magnetic helicity of the current-carrying part of the magnetic field, and $\hpj$ is the volume threading helicity between $\Bpvec$ and the current-carrying field. Because $\Bvec$ and $\Bpvec$ are designed such that they share the same normal distribution on $\partial\mathcal{V}$, not only $\hv$, but also both $\hj$ and $\hpj$ are independently gauge invariant. 

In contrast to $\hv$, however,  $\hj$ and $\hpj$ are not conserved in ideal MHD, as shown recently by \cite{2018ApJ...865...52L}, who provided the first analytical derivation of the time variation of these helicities. From their analytical study and their analysis of different numerical experiments, they revealed the existence and key role of a gauge-invariant transfer term between $\hj$ and $\hpj$, that enables the exchange between the different contributions to $\hv$.

The properties of $\hj$ and $\hpj$ have been investigated in a few works only so far. \cite{2014SoPh..289.4453M} studied them, based on three-dimensional MHD models of the emergence of a twisted magnetic flux tube, that resulted in the formation of a small active region (AR) in the model corona. Two experiments have been analyzed, a ``non-eruptive'' and an ``eruptive'' one. In the eruptive case, part of the model magnetic structure is ejected from the simulation volume at least once during the simulation time span, while in the non-eruptive case the magnetic field remains confined within the model volume. It was found that at least $\hj$ showed pronounced fluctuations around the onset of the model mass ejection in the eruptive simulations.

\cite{2017A&A...601A.125P} presented a study based on seven different three-dimensional visco-resistive MHD simulations of the emergence of a twisted model flux rope into a stratified model atmosphere, that resulted in either a stable (non-eruptive) or an unstable (eruptive) coronal configuration. While the basic setup in all of these simulations was identical, only the strength and direction of the background (surrounding) magnetic field was modified to obtain the different solutions. They concluded that, for the analyzed set of numerical experiments $\hv$ clearly discriminated between stable and unstable simulations, in contrast to, \eg, total, potential, and free magnetic energy, as well as magnetic flux. A generally higher $\hv$ in the stable simulations, however, disqualified $\hv$ as a useful quantity to predict eruptive behavior, at least in cases where the self and mutual helicities are of opposite sign \citep[see also, \eg,][]{2005ApJ...624L.129P}. 

In contrast, significantly greater values of $\hj$ during the pre-eruptive phase, and especially during the time of strong flux emergence, were noticed from the unstable simulations studied in \cite{2017A&A...601A.125P}. Even more powerful, the ratio of the current-carrying to total helicity, $|\hj|/|\hv|$, turned out as to represent a most fruitful proxy for eruptivity, with values $\gtrsim0.45$ prior to the model eruptions, in contrast to the corresponding value for the stable (non-eruptive) configuration. As noted by the authors, the threshold $|\hj|/|\hv|\simeq0.45$ is not to be regarded as a universal one, but rather depends on the properties of the particular analyzed case.

In another recent study, \cite{2018ApJ...863...41Z} investigated the helicity-based eruptivity threshold using three-dimensional line-tied MHD simulations, in which eruptivity was imposed by controlled motions, driven on the lower boundary of the simulation domain. These motions were designed such as to mimic the long-term evolution of solar ARs, including shearing motions and magnetic diffusion on large scales. Starting from the same initial field configuration that contained a flux rope, the different numerical simulations were based on different types of boundary motions that led to the eruptive evolution. The authors noted a value of $|\hj|/|\hv|\simeq0.3$ at the onset times of torus instability, for all simulations, \ie, independently of how the system was destabilized.

As a side result, analyzing a simulation of the generation of a solar coronal jet, \cite{2018ApJ...865...52L} also found that the jet was triggered for large values of the ratio $|\hj|/|\hv|$, though the focus of the study was primarily on the analysis of the properties of $\hj$ and $\hpj$.
 
So far, only few works attempted to investigate the decomposed helicity for observational cases. In \cite{2018ApJ...855L..16J}, a nonlinear force-free (NLFF) model of AR~11504 one hour prior to a filament eruption was used to calculate the contributions to the total helicity. They found  $|\hj|/|\hv|=0.17$, underlying that the thresholds for eruptivity given in \cite{2017A&A...601A.125P} and \cite{2018ApJ...863...41Z} are valid with regards to the particular analyzed simulations only. 

\cite{2014SoPh..289.4453M} was the first to attempt the monitoring of the long-term evolution of the individual contributors to magnetic helicity for two solar ARs (11072 and 11158, prolific in confined and eruptive flaring, respectively). The time evolution of $\hj$  showed a clear correspondence to rapid flux emergence and the formation of a filament and a X2.2 flare in AR~11158, despite the rather low time cadence of the underlying NLFF models (four hours). The corresponding analysis of AR~11072 was hampered by a non-satisfactory level of solenoidality of the underlying NLFF solutions. 

Just recently, two studies dealt with the long-term evolution of the magnetic energy and helicity budgets in solar ARs that hosted major flares, based on helicity computations of unprecedented accuracy, within the application to observed data. \cite{2019arXiv190706365M} analyzed the helicity and energy budgets of AR~12673, in the course of two major flares (a preceding confined and a following eruptive X-flare). They found distinct local maxima in time evolution of $|\hj|/|\hv|$ allowing them to suggest an approximate threshold of $|\hj|/|\hv|\simeq0.15$ for the eruptivity in that AR. Their results are in line with that of \cite{2019ApJ...880L...6T} who analyzed the coronal evolution of AR~11158. Though the primary focus of the latter study was on the sensitivity of the magnetic helicity computation with respect to the solenoidal property of the underlying NLFF solution (for more details see also \href{sss:mf_modeling}{Sect.~\ref{sss:mf_modeling}} in the present study), an approximate characteristic pre-flare level of $|\hj|/|\hv|\simeq0.2$ can be identified in their Fig.~4(d).

In our work, we go further and provide the first study of the (decomposed) magnetic helicity budget in two solar ARs of different respective flare profile (prolific in either major confined or major eruptive flares) and evolutionary stage (well-developed vs. fast evolving with rapid flux emergence). For this purpose, we study AR~11158 during February~2011 and AR~12192 during October~2014, respectively. We analyze the coronal magnetic helicities and energies in the course of confined and eruptive flaring, to study their potential to discriminate the two types of flaring timely before their occurrence. Importantly, we base our analysis on optimized NLFF time series, with highly satisfactory force-free and solenoidal properties, to allow helicity computations of unprecedented accuracy. Moreover, we incorporate the results of three different helicity computation methods, in order to explore the possible spread of values obtained.

\section{Methods}\label{sec:methods}

\subsection{AR selection}\label{subsec:selection}

We aim to compare the coronal magnetic energy and helicity of two solar ARs, prolific in major ({\it GOES} class M5.0 and larger) eruptive and confined flares. AR AR~11158, produced the first X-class flares of solar cycle 24. All major flares of this AR, as observed during disk passage in February~2011, were associated with CMEs. In contrast, AR AR~12192 showed a flare profile that clearly deviates from known flare-CME statistics \citep[\eg,][]{2006ApJ...650L.143Y} in that, during its disk passage in October 2014, it produced six confined X-class flares, but none was associated with a CME.

\subsection{Data}
\label{subsec:data}

The magnetic characteristics were studied based on photospheric vector magnetic field data \citep{2014SoPh..289.3483H}, derived from {\it Solar Dynamics Observatory} \citep[\sdo;][]{2012SoPh..275....3P} Helioseismic and Magnetic Imager \citep[HMI;][]{2012SoPh..275..229S} polarization measurements. In particular, the {\sc hmi.sharp\_cea\_720s} data series was used which contains a Lambert Cylindrical Equal-Area projected magnetic field vector, decomposed into $\Br$, $\Bphi$, and $\Btheta$ at each remapped grid point, within automatically identified active-region patches \citep[][]{2014SoPh..289.3549B}. These spherical components relate to the heliographic magnetic field components, as defined in \cite{1990SoPh..126...21G}, as $(\Bx,\By,\Bz)=(\Bphi,-\Btheta,\Br)$, where $x$, $y$ and $z$ denote the solar positive-westward, positive northward and vertical direction, respectively. The native resolution of the photospheric field data is 0.03 CEA-degree, corresponding to $\approx$~360~km\,pixel$^{-1}$ at disk center. 

Within large sunspot umbrae, unreasonable magnetic field values with high errors are sometimes present in HMI data products. The center of the negative-polarity sunspot in AR~12192 represents such a case, with a patch of abnormally weak $B_z$ (i.e., $B_r$). In order to compensate for the artificial magnetic profile within the sunspot umbra, we use the irregularly sampled but known and accurately measured magnetic field values, to interpolate smoothly over the grid of erroneous measurements, using bilinear interpolation.

\subsection{Modeling}
\label{subsec:modeling}

\subsubsection{Magnetic field modeling}
\label{sss:mf_modeling}

For NLFF modeling, we binned the photospheric data by a factor of four, to a resolution of $\approx$~2\asecs\,pixel$^{-1}$, while almost preserving the magnetic flux. The adopted computational domains are of the extent of $148\times92\times128$~pixel$^3$ and $276\times200\times128$~pixel$^3$, to model the force-free corona of AR~11158 and 12192, respectively. The NLFF equilibria are computed using the method of \cite{2010A&A...516A.107W}. In this way, we obtain the quasi-static evolution of the solar corona in and around AR~11158 from 2011 February~11 19:00~UT to February~15 23:59~UT, and for AR~12192 from 2014 October~ 20 06:59~UT to October~25 11:59~UT. Around the time of intense flares (equal or larger {\it GOES} class M5.0), we use the native time cadence of 12 minutes and use an 1-hour cadence otherwise.

Two controlling parameter are frequently used to quantify the consistency of the obtained NLFF solutions. The current-weighted angle between the modeled magnetic field and electric current density, $\thetaj$, \citep[][]{2006SoPh..235..161S} and the volume-averaged fractional flux, $\fiavg$, \citep[][]{2000ApJ...540.1150W}, which is a measure of local deviations from solenoidality within the model volume. As a rule of thumb, the smaller the corresponding values are, the more force- and divergence-free a NLFF solution is. For a perfectly force-free and solenoidal NLFF solution, $\thetaj=0$ and $\fiavg=0$. 

For the NLFF time series of AR~11158, we find median values of $\thetaj=15.6\dc\pm2.7\dc$ and $\fiavg\times10^{4}=2.23\pm0.98$. The corresponding estimates for AR~12192 are $\thetaj=5.62\dc\pm0.16\dc$ and $\fiavg\times10^{4}=3.46\pm0.25$, underlying the high quality of the NLFF fields for subsequent reliable helicity computation \citep[see][for a dedicated study]{2019ApJ...880L...6T}.

Also \cite{2016SSRv..201..147V} highlighted that, in order to guarantee a reliable computation of magnetic helicity, the input magnetic field has to fulfill certain requirements concerning its divergence-freeness, \ie, how well $\nabla\cdot\Bvec$ is satisfied. It was shown that if the ratio $\Ediv/\Etot\gtrsim0.1$, the error in the computation of $\hv$ may grow considerably (see their Sect.~7 and Fig.~8(b)). The expression $\Ediv$ is based on the decomposition of the magnetic energy within $\mathcal{V}$ by \cite{2013A&A...553A..38V}, in the form
\begin{eqnarray}
\Etot&=&\frac{1}{2\mu_0}\int_\mathcal{V} B^2 {\rm ~d}\mathcal{V} = \Epot + \Ej \nonumber \\
&=&\Eps+\Ejs+\Epns+\Ejns+\Emix, \label{eq:e_i}
\end{eqnarray}
with $\Epot$ and $\Ej$ being the energies of the potential and current-carrying magnetic field, respectively. $\Epot$ is used to compute an upper limit for the free energy as $\Efree=\Etot-\Epot$. $\Eps$ and $\Ejs$ are the energies of the potential and current-carrying solenoidal magnetic field components. $\Epns$ and $\Ejns$ are those of the corresponding non-solenoidal components. All terms are positive-defined, except for $\Emix$, which corresponds to all cross terms \citep[see Eq.~(8) in][for the detailed expressions]{2013A&A...553A..38V}. For a perfectly solenoidal field, one finds $\Eps=\Epot$, $\Ejs=\Ej$, and $\Epns=\Ejns=\Emix=0$. Based on \href{eq:e_i}{Eq.~(\ref{eq:e_i})}, the divergence-based energy contribution is defined as $\Ediv=\Epns+\Ejns+|\Emix|$. 

Recently, an extension to the work of \cite{2016SSRv..201..147V} who investigated the corresponding effect based on an idealized model, was presented by \cite{2019ApJ...880L...6T} who considered the dependency of  helicity computations on the field's solenoidal property in NLFF time series based on solar observations.  It was shown that helicity computations may be meaningful and trustworthy only, if $\Ediv/\Etot\lesssim0.05$ and $\fiavg\times10^4\lesssim5$, for the underlying coronal magnetic field model. In the present work, both NLFF time series have a median value of $\Ediv/\Etot\lesssim0.01$, and thus, we may safely assume a correspondingly small error in the helicity computations. Note that these values, together with that of $\fiavg$ listed above, are considerably better than in earlier works \citep[\eg,][]{2014SoPh..289.4453M,2018ApJ...855L..16J,2019arXiv190706365M}.

\subsubsection{Magnetic helicity computation}
\label{sss:mh_modeling}

The 3D cubes containing the NLFF magnetic field are used as an input to three different finite-volume (FV) helicity computation methods. In brief, the method of \cite{2011SoPh..272..243T} to compute the relative helicity, solves systems of partial differential equations to obtain the vector potentials $\Avec$ and $\Apvec$, employing the Coulomb gauge, $\nabla\cdot\Avec=\nabla\cdot\Apvec=0$, \citep[``\FVjt'' hereafter; see also Sect.~2.1 of][for details]{2016SSRv..201..147V}. The methods of \cite{2012SoPh..278..347V} and \cite{2014SoPh..289.4453M} are based on an integral formulation for the vector potentials within a finite volume and employ a DeVore gauge, $A_z=A_{{\rm p},z}=0$, \citep[see also Sect.~2.2 of][]{2016SSRv..201..147V}. The two methods differ in the way in which the Laplace equation for the potential field solution is solved numerically, as well as the numerical calculation of the involved integrals and derivatives. The method of \cite{2012SoPh..278..347V} is referred to as ``\FVgv'', hereafter. We consider two realizations of the \FVkm\ method, one where the rectangle integration rule is used (``\FVkm'') , and one where the weighted trapezoidal rule is used (``\FVkmw'') to compute the involved integrals. All methods define the reference field as $\Bpvec=\nabla\varphi$, with $\varphi$ being the scalar potential, subject to the constraint $\nabla_n\varphi=\Bvec_n$ on $\partial\mathcal{V}$, and solve the corresponding Laplace equation for $\varphi$ with different methods.
 
The methods have been tested in the framework of an extended proof-of-concept study on finite-volume helicity computation methods \citep[][]{2016SSRv..201..147V}, where it has been shown that for various test setups the methods deliver helicity values in line with each other, differing by a few percent only, given a sufficiently low level of $\nabla\cdot\Bvec$ in the underlying magnetic test case. 

\subsection{Analyzed quantities}

We apply the different FV helicity computation methods to the two time series of NLFF extrapolations for AR~11158 and AR~12192, and correspondingly obtain four, possibly differing, results for each time instant for the extensive quantities $\hv$ and its contributors, $\hpj$ and $\hj$, as well as for $\Epot$, and thus $\Efree$, since the different FV helicity computation methods derive $\Bpvec$ in a different numerical way. For each time instance, we compute the mean values $\bar\Epot$, $\bar\Efree$, $\bar\hv$, $\bar\hj$, and $\bar\hpj$, and consider this as to be the most representative approximation of the real values.

In order to make the computed quantities of the two different ARs better comparable, we also calculate intensive quantities. We define the normalized helicity as $\hvprime$, where $\phi=\frac{1}{2}\int_{\mathcal{S}(z=0)} |B_z| d\mathrm{S}$, \ie, half of the total unsigned magnetic flux, $\phi$, across the NLFF lower boundary. Often employed proxies to quantify the non-potentiality and eruptivity of the considered magnetic configuration are the form of the free energy ratio, $\efprime$, where $\Etot$ is the total magnetic energy of the input NLFF fields, and the helicity ratio, $\hjprime$.

All mean values are presented and interpreted in context with the spread of the four values obtained for the individual physical quantities, where we define the spread as to be bounded by the two estimates which deviate the most from the respective mean value.

\section{Results}
\label{sec:results}

\subsection{Coronal magnetic field structure}
\label{subsec:nlff_mf_config}

\begin{figure}[t]
	\centering
	\includegraphics[width=0.45\textwidth]{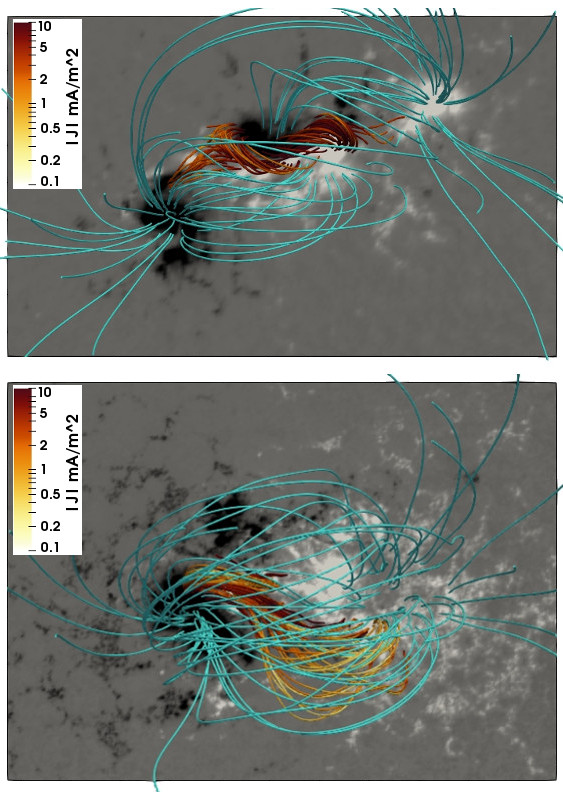}
	\put(-190,310){\color{white}\bf(a)}
	\put(-190,155){\color{white}\bf(b)}
\caption{NLFF magnetic field of (a) AR~11158 on 2011 February~14 at 21:00~UT and (b) AR~12192 on 2014 October~24 at 19:00~UT. Field lines outlining the large-scale magnetic field are colored turquoise. Sample field lines in the centers of the respective ARs are color-coded according to the magnitude of the total current density, $|\vec{J}|$. The gray scale background shows the vertical magnetic field component at a photospheric level, saturated at $\pm\,1$~kG.
}
\label{fig:nlff_fls}
\end{figure}

In agreement with previous works, we find the coronal magnetic field above AR~11158 on February~14 at $\sim$21:00~UT in the form of a low-lying magnetic flux rope aligned with the main polarity inversion line and surrounded by the large-scale field associated to the strong westernmost positive and easternmost negative polarity patches of the AR (\href{fig:nlff_fls}{Fig.~\ref{fig:nlff_fls}(a)}). Similar model results have been presented and discussed in, \eg, \cite{2012ApJ...752L...9J,2012ApJ...748...77S,2013ApJ...770...79I}.

The coronal magnetic field configuration above AR~12192 on October~24 at $\sim$19:00~UT appears in the form of a low-lying weakly twisted flux rope above the main polarity inversion line of the AR (\href{fig:nlff_fls}{Fig.~\ref{fig:nlff_fls}(b)}), bridged by the large-scale magnetic field associated to the dispersed magnetic field surrounding. Similar model results have been presented in, \eg, \cite{2015RAA....15.1537J,2015ApJ...804L..28S,2016ApJ...818..168I}. 

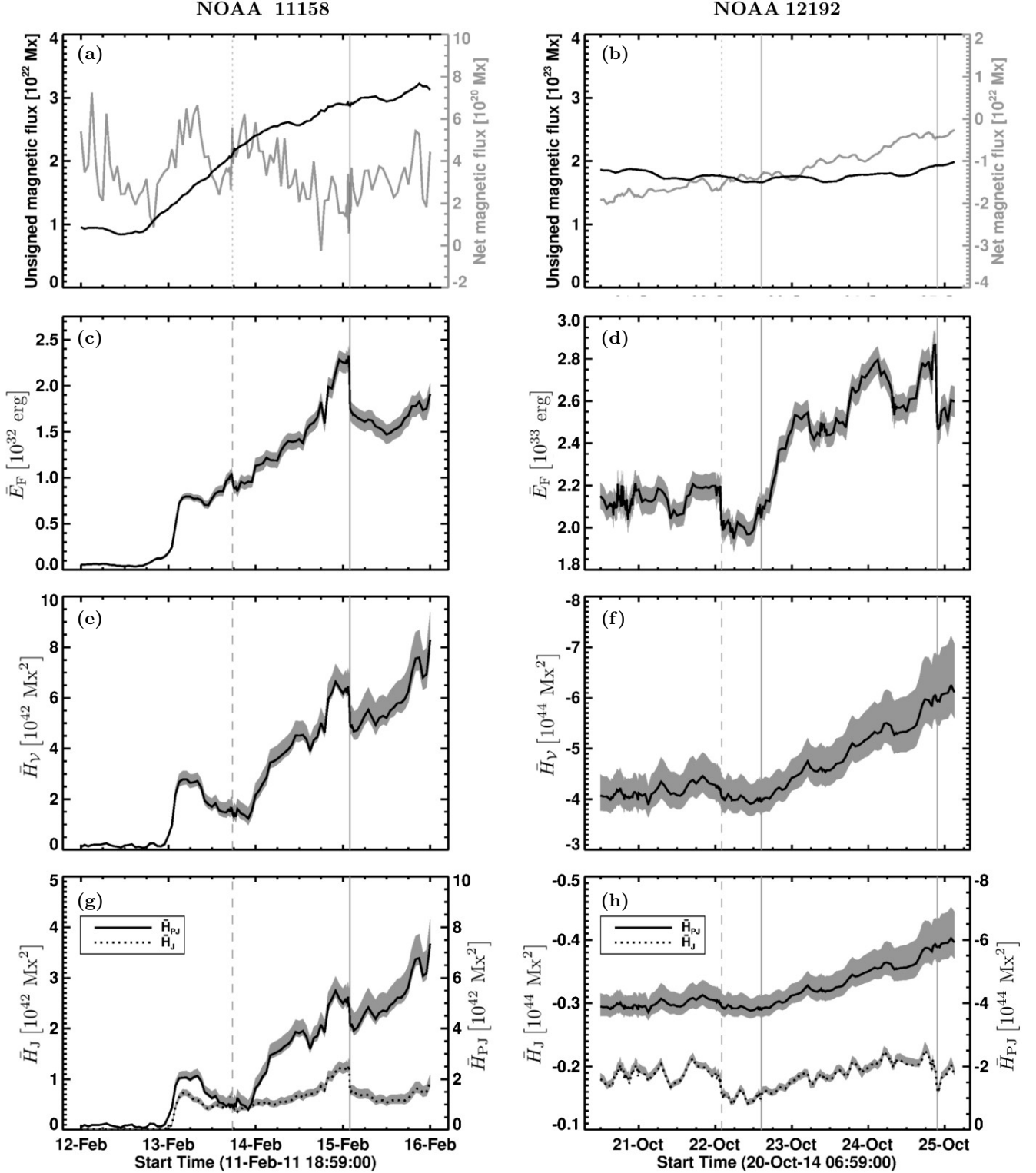
\begin{figure*}
	\centering
	\includegraphics[width=\textwidth]{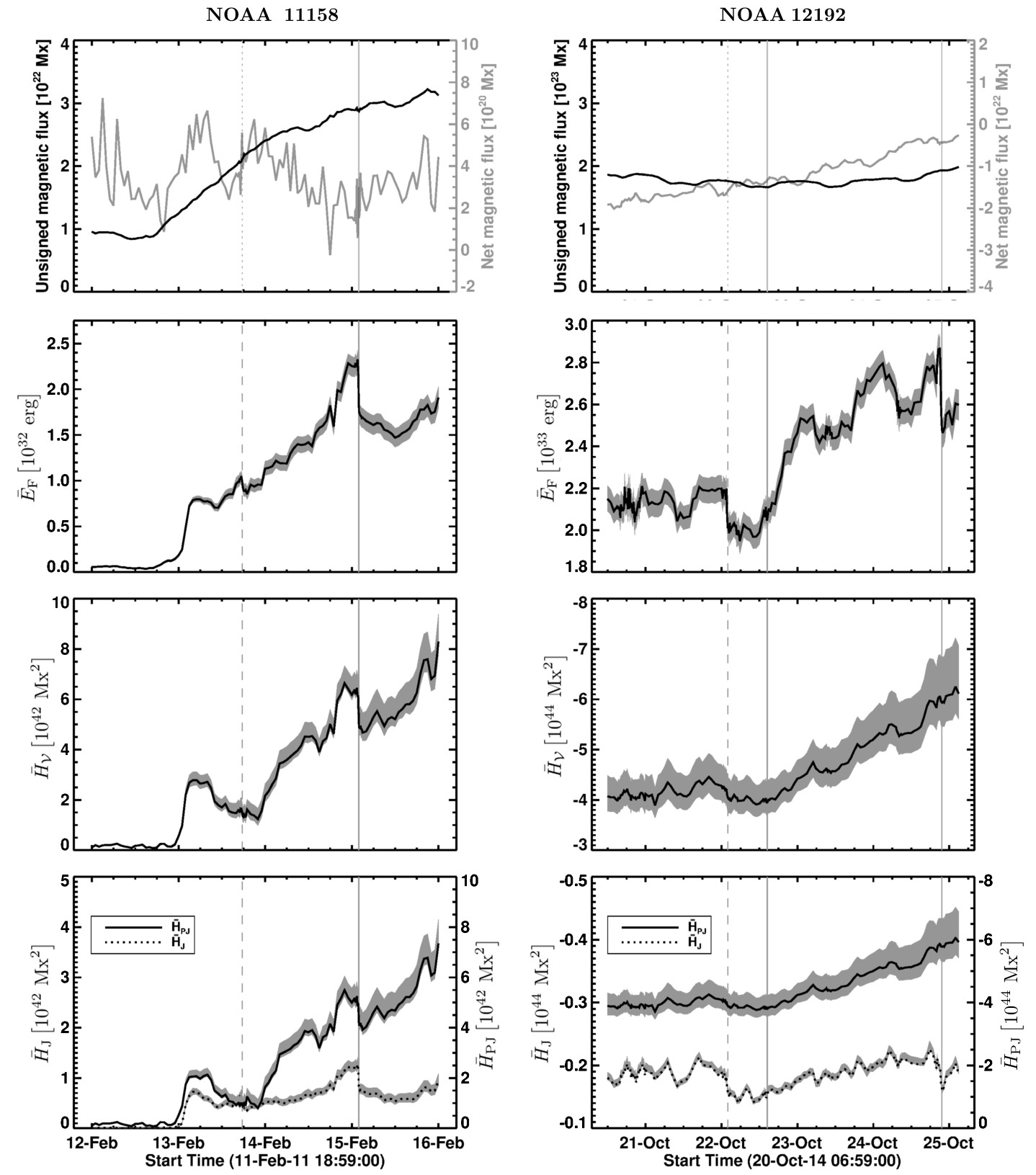}
	\put(-470,550){\bf(a)}
	\put(-212,550){\bf(b)}
	\put(-470,410){\bf(c)}
	\put(-212,410){\bf(d)}
	\put(-470,272){\bf(e)}
	\put(-212,272){\bf(f)}
	\put(-470,133){\bf(g)}
	\put(-212,133){\bf(h)}
\caption{
Time evolution of different extensive quantities for AR~11158 (left column) and AR~12192 (right column). The net magnetic flux (gray curve) and total unsigned flux ($\phi$; black curve) are shown in (a) and (b), the mean free magnetic energy, $\bar\Efree$, in (c) and (d), and the mean magnetic helicity, $\bar\hv$, in (e) and (f), respectively. The contributions of the current-carrying ($\bar\hj$; dotted line) and volume-threading ($\bar\hpj$; solid line) helicty are shown in (g) and (h), respectively. Black curves in (c)--(f) represent the mean values of the quantities computed with the different FV methods. The shaded areas represent the spreads of the respective quantities, bounded by those which lie farthest away from the mean value. Vertical dashed and solid lines mark the {\it GOES} peak time of M- and X-class flares, respectively.
}
\label{fig:primary}
\end{figure*}

\subsection{Extensive quantities: \\ Magnetic flux, free magnetic energy, and helicities}
\label{sss:extensive}

\subsubsection{AR~11158}

Upon emergence, AR~11158 exhibited a considerable increase in the total unsigned flux (\href{fig:primary}{Fig.~\ref{fig:primary}(a)}; black line), starting from late February~12, which corresponds to the time when a pronounced filament was emerging, as analyzed in detail by, \eg, \cite{2012ApJ...748...77S}. Parts of the filament erupted during two eruptive flares, an M6.6 flare (SOL2011-02-13T17:38) and an X2.2 flare (SOL2011-02-15T01:56).

The corresponding evolution of the mean free magnetic energy, $\bar\Efree$, is shown in \href{fig:primary}{Fig.~\ref{fig:primary}(c)}. The spread of the energy values deduced from the different FV methods is shown as gray shaded area, and is bounded by the results from the \FVkm/\FVjt\ method at higher/lower energies. Two distinct episodes can be distinguished. First, a considerable increase of $\bar\Efree$, co-temporal with the strong flux emergence, resulting in a free magnetic energy of $\approx0.8\times10^{32}~\erg$ early on February~13. Second, notable decreases of $\bar\Efree$ are observed around the two major eruptive flares (the M6.6 and X2.2 flare, marked by a vertical dashed and solid line, respectively). The trends just discussed compare well with results previously published in literature \citep[\eg,][]{2012ApJ...748...77S,2015RAA....15.1537J,2013ApJ...772..115T}. 

The time evolution of the mean magnetic helicity, $\bar\hv$ (\href{fig:primary}{Fig.~\ref{fig:primary}(e)}), also reflects the emergence of the magnetic flux rope, with $\bar\hv$ increasing from $\approx0.1\times10^{42}\mxmx$ to $\approx3\times10^{42}\mxmx$. The response to the occurring eruptive flares is reflected by a response similar to that of $\bar\Efree$, with a (smaller) larger decrease during the (M6.6) X2.2 flare. These trends are in overall agreement with the results presented by \cite{2012ApJ...752L...9J,2015RAA....15.1537J}.

The individual contributions of the current-carrying ($\bar\hj$) and volume-threading ($\bar\hpj$) helicities are shown in \href{fig:primary}{Fig.~\ref{fig:primary}(g)} (dotted and solid curve, respectively). Throughout the considered time period, both, $\bar\hj$ and $\bar\hpj$ are positive, with $\bar\hpj$ being the dominated contributor to $\bar\hv$ at most time instances (being a factor of 2--10 larger than $\bar\hj$). 

\subsubsection{AR~12192}

AR~12192, the largest solar AR observed during the past $\sim$24 years, exhibited a more or less constant and unusually high unsigned magnetic flux ($\approx2\times10^{23}$~Mx) during disk passage \citep[see also, \eg, Table~1 of][]{2015ApJ...804L..28S}. The little variation resulted from the slow time evolution of the well-developed AR, in absence of strong flux emergence (\href{fig:primary}{Fig.~\ref{fig:primary}(b)}). 

\begin{figure*}
	\centering
	\includegraphics[width=\textwidth]{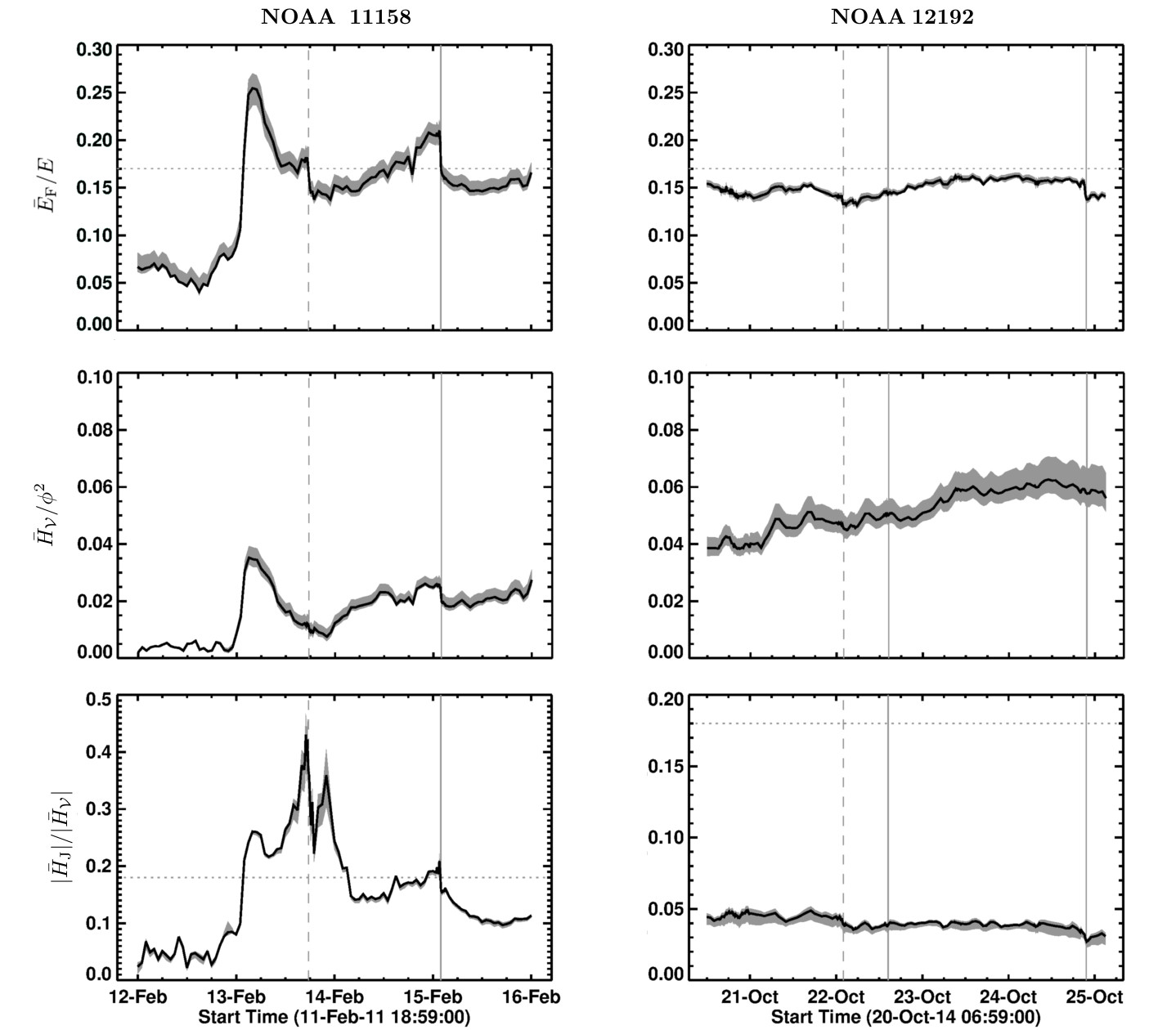}
	\put(-455,415){\bf(a)}
	\put(-202,415){\bf(b)}
	\put(-455,272){\bf(c)}
	\put(-202,272){\bf(d)}
	\put(-455,132){\bf(e)}
	\put(-202,132){\bf(f)}
\caption{
Time evolution of different intensive quantities for AR~11158 (left panels) and AR~12192 (right panels). The magnetic energy ratio, $\efprime$, is shown in (a) and (b), the respective normalized helicity, $\hvprime$, in (c) and (d), and the helicity ratio, $\hjprime$, in (e) and (f), respectively. Black curves represent the mean values of the quantities computed with the different FV methods. The shaded areas represent the spreads of the respective quantities, bounded by those which lie farthest away from the mean value. Vertical dashed and solid lines mark the {\it GOES} peak time of M- and X-class flares, respectively.\\
}
\label{fig:proxies}
\end{figure*}
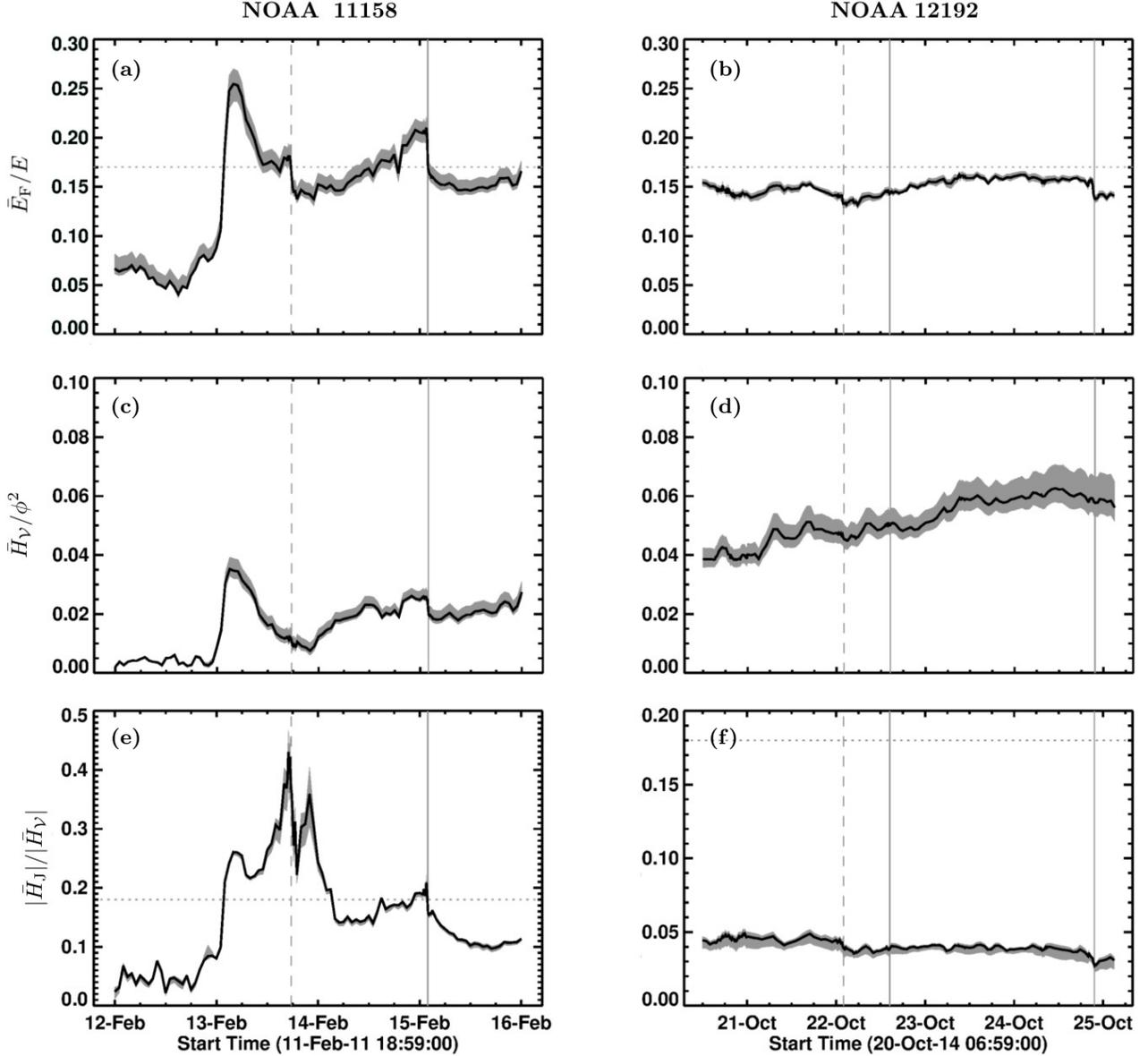

The corresponding evolution of $\bar\Efree$ (\href{fig:primary}{Fig.~\ref{fig:primary}(d)}) is characterized by distinct variations around three major confined flares (SOL2014-10-22T01:59M8.7, 2014-10-22T14:28X1.6, and 2014-10-24T21:41X3.1), with the spread of solutions being bound by that of the \FVkm\ and \FVjt\ method at higher and lower energies, respectively. Despite the similar trend, we find $\bar\Efree$ by a factor of 10 higher than \cite{2015RAA....15.1537J}, and in the approximate range 2--$3\times10^{33}~\erg$. Given the unusually high unsigned magnetic flux, we regard our numbers as highly plausible, however. This is further substantiated by an estimated mean free magnetic energy of $\approx$15\%, which for an AR with a well defined flux rope is highly realistic (see \href{sss:intensive_12192}{Sect.~\ref{sss:intensive_12192}} and \href{fig:proxies}{Fig.~\ref{fig:proxies}(b))}.

$\bar\hv$ was negative increasing from about $-4\times10^{44}\mxmx$ to $-6\times10^{44}\mxmx$ during the considered time period (see \href{fig:primary}{Fig.~\ref{fig:primary}(f)} and note the reversed $y$-axis labeling). The time evolution of $\bar\hv$ shows hardly any sensitivity towards the occurrence of the major confined flares. The spread of the solutions of the individual methods is bounded by the values derived from the \FVjt\ (\FVgv) method at higher (lower) values. Similar as for $\bar\Efree$ before, also our estimates of $\bar\hv$ are by a factor of 10 larger than those presented in \cite{2015RAA....15.1537J}, still showing a similar trend. Yet again, our results are compatible with the strong magnetic flux in AR~12192, substantially higher than that of ``typical'' ARs \citep[a few $10^{22}\mx$; see, \eg, Fig.~4 of][]{2017ApJ...834...56T}. Also, our results are highly reliable, given the low value of $\Ediv/\Etot$ in the underlying NLFF models (see \href{sss:mf_modeling}{Sect.~\ref{sss:mf_modeling}}). 

$\bar\hv$ is dominated by the contribution of $\bar\hpj$ at all times (\href{fig:primary}{Fig.~\ref{fig:primary}(h)}; solid line), with $\bar\hj$ being by a factor of $\sim$25 smaller (\href{fig:primary}{Fig.~\ref{fig:primary}(h)}; dotted line). The spread of solutions for $\hpj$ is bounded by that of \FVjt\ and \FVgv\ at high and low helicities. That of $\hj$ is bound by \FVkm\ and \FVgv, respectively.

\subsection{Intensive quantities: \\ Normalized helicity and eruptivity proxies}
\label{sss:intensive}

The goal of our study is to compare ARs that hosted almost exclusively major confined (AR~12192) or eruptive flares (AR~11158), during their disk passage. The analysis of the extensive quantities $\bar\Efree$, $\bar\hv$, $\bar\hj$, and $\bar\hpj$, above (see \href{sss:extensive}{Sect.~\ref{sss:extensive}}), revealed quite some differences between the two ARs. Well-developed and slowly evolving AR~12192 hosted a total unsigned magnetic flux, $\phi$, and free magnetic energy, $\bar\Efree$, and a helicity of the current-carrying field, $\bar\hj$, about 10 times larger than the newly formed and rapidly evolving AR~11158. Only, the decomposition of $\bar\hv$ into $\bar\hj$ and $\bar\hpj$, revealed that $\bar\hv$ in AR~12192 exceeded that of AR~11158 by a factor of 100, due to the contribution of the volume-threading helicity, $\bar\hpj$. In order to more easily compare the two different ARs, we analyze intensive quantities in the following.

\subsubsection{AR~11158}

The energy ratio shows increasing trends prior to the eruptive flares and values $\efprime\gtrsim0.17$ (\href{fig:proxies}{Fig.~\ref{fig:proxies}(a)}; see horizontal dashed line for reference). Highest values are obtained for the time period related to the strong flux emergence, with $\efprime\simeq0.25$. Prior to the presence of strong magnetic fluxes (before late February~12), it is considerably smaller ($\efprime\lesssim0.1$). The spread of values of $\efprime$ is bounded by the solutions of \FVkm\ and \FVjt\ at its higher and lower bound, respectively.

Peak values of $\hvprime\gtrsim0.05$ are also found around the time of strong flux emergence (early on February~13), while it is $\lesssim0.04$ at most other times (\href{fig:proxies}{Fig.~\ref{fig:proxies}c}). The spread of solutions is bound by the results obtained with the \FVkm\ and \FVjt\ at higher and lower values, respectively. 

$\hjprime$ shows an (in-) decreasing trend (before) after the major eruptive flares, with pre-flare values $\hjprime\gtrsim0.17$ (\href{fig:proxies}{Fig.~\ref{fig:proxies}(e)}; see horizontal dashed line for reference). Note also the little spread of the results based on the different methods. The spread of solutions is bound by the results obtained with the \FVkm\ and \FVjt\ at higher and lower values, respectively.  

\subsubsection{AR~12192}
\label{sss:intensive_12192}

The energy ratio shows no clear trends prior to the occurrence of the major confined flares, and values $\efprime\lesssim0.17$ at all times (\href{fig:proxies}{Fig.~\ref{fig:proxies}(b)}; see horizontal dashed line for reference). Correspondingly, no common characteristic pre-flare level in context with the major flares can be identified. The spread of values is bounded by the solutions of \FVkmw\ and \FVjt\ at its higher and lower bound, respectively. 

$\hvprime$ shows a smooth and slowly increasing trend, with values $\hvprime\gtrsim0.05$ at most times (\href{fig:proxies}{Fig.~\ref{fig:proxies}(d)}). The spread is bounded by the solutions of \FVjt\ (\FVgv) at high (low) values.

Very little variation of $\hjprime$ is found around the time of the major confined flares, including no significant increase, or a characteristic pre-flare value (\href{fig:proxies}{Fig.~\ref{fig:proxies}(f)}). The spread is bounded by the solutions of \FVjt\ (\FVgv) at low (high) values.

\section{Discussion and Summary} 
      \label{s:discussion}

We aimed to compare the coronal magnetic energy and helicity of two solar ARs, prolific in major eruptive (AR~11158) and confined (AR~12192) flares, and analyze the potential of proxies for eruptivity ($\efprime$, $\hjprime$) to hint at the upcoming flares. AR~11158 was rapidly evolving and produced the first major flares of solar cycle 24, all associated with CMEs. In contrast, well-developed and slowly evolving AR~12192 produced six major confined X-class flares (\ie, no associated CMEs). 

Our results are based on the application of three different numerical approaches to compute the relative helicity \citep{2011SoPh..272..243T,2012SoPh..278..347V,2014SoPh..289.4453M}. Compared to previous works, we based our energy and helicity computations on time series of NLFF model solutions with unprecedented quality regarding their fulfillment of the solenoidal condition ($\fiavg\times10^4\lesssim4$ and $\Ediv/\Etot\lesssim0.01$), supporting the high reliability of our main findings:

\begin{itemize}[leftmargin=16pt]
\item[\sc{(i)}] For both ARs, $\bar\hv$ and $\bar\Efree$ exhibit a similar time evolution (\href{fig:primary}{Fig.~\ref{fig:primary}(c)}--\href{fig:primary}{\ref{fig:primary}(f)}). Timely centered around the emergence of strong magnetic flux, as well as the occurrence of major flares, we detect significant changes only for CME-productive AR~11158. 
\item[\sc{(ii)}] For the analyzed ARs, $\bar\hv$ was dominated by the contribution of $\bar\hpj$ (\href{fig:primary}{Fig.~\ref{fig:primary}(g)}, \href{fig:primary}{\ref{fig:primary}(h)}). Noteworthy, while the absolute value of $\hpj$ exceeds that of $\hj$ by a factor of 2--10  in AR~11158, it is about $\sim 25$ times larger in AR~12192, probably due to the unusually large unsigned magnetic flux ($\phi\propto10^{23}$~Mx).
\item[\sc{(iii)}] On average, $\hvprime$ is larger for AR~12192 ($\gtrsim0.05$; \href{fig:proxies}{Fig.~\ref{fig:proxies}(d)}), compared to that of AR~11158 ($\lesssim0.04$;  \href{fig:proxies}{Fig.~\ref{fig:proxies}(c)}). The noteworthy exception is the period of strong flux emergence early on February~13, with peak values $\hvprime\gtrsim0.13$ for AR~11158. 
\item[\sc{(iv)}] While the eruptivity proxy, $\hjprime$, increases strongly before major eruptive flares in AR~11158, only little variation is found for AR~12192 (\href{fig:proxies}{Fig.~\ref{fig:proxies}(e)} and \href{fig:proxies}{\ref{fig:proxies}(f)}, respectively). A corresponding statement holds for $\efprime$ (\href{fig:proxies}{Fig.~\ref{fig:proxies}(a)} and \href{fig:proxies}{\ref{fig:proxies}(b)}, respectively). For both, $\efprime$ and $\hjprime$, characteristic pre-flare values in AR~11158 are $\gtrsim0.17$.
\item[\sc{(v)}] $\hjprime$ does not scale with the size of the flares in NOAA~11158. We find values of  $\hjprime\gtrsim0.4$ ($\gtrsim0.17$) prior to the eruptive M6.6 (X2.2) flare, respectively (see \href{fig:proxies}{Fig.~\ref{fig:proxies}(e)}).

\item[\sc{(vi)}] A pronounced response of $\hjprime$ on the occurrence of flares is only seen for the major eruptive flares (\ie, for AR~11158; see \href{fig:proxies}{Fig.~\ref{fig:proxies}(e)}). 
\end{itemize}
 
In summary, our findings substantiate the suggestion of \cite{2017A&A...601A.125P} that the helicity ratio $|\hj|/|\hv|$ shows a strong ability to indicate the eruptive potential of a magnetic system, and that peak values are to be expected prior to eruptive flaring. Our results also support the findings of \cite{2018ApJ...863...41Z} and \cite{2018ApJ...865...52L} in that a close correlation may exist between large values of the helicity ratio and eruptivity. In our work, these findings are based on real solar observations of two different ARs, whereas the aforementioned studies were based on numerical simulations. 

The analysis of \cite{2017A&A...601A.125P} was based on numerical simulations of a solar-like AR, that involved distinct reorganizations of the model coronal magnetic field. More precisely, a flux rope rises from the convection zone to reconnect with the magnetic field in the low atmosphere above, to form a secondary twisted flux rope. This secondary flux rope is either stable \citep[in the non-eruptive simulations;][]{2013ApJ...778...99L} or unstable \citep[in the eruptive simulations;][]{2014ApJ...787...46L}. We may therefore compare our helicity analysis of AR~12192 and 11158 with the corresponding analysis of the stable and unstable simulations, respectively, by \cite{2017A&A...601A.125P}.

\cite{2017A&A...601A.125P} suggested that $|\hj|/|\hv|$ is (smaller) larger in (non-) eruptive cases, based on numerical simulations, composed of a model flux rope emerging into an overlying arcade field (non-) favorable for magnetic reconnection. For a given dipole strength of the overlying field arcade, they found $|\hv|$ to be (larger) smaller for the (non-) eruptive case (compare, \eg, violet dash-dotted and red dashed lines in their Fig.~5(a)), if the orientation of the upper part of the poloidal field of the flux rope was oriented parallel (anti-parallel) with respect to the overlying arcade field. 

Similarly, In our work, we find (smaller) larger values of $\hjprime$ for (non-) eruptive AR (12192) 11158. We assume that the smaller $\hjprime$ in AR~12192 can be attributed to the substantially higher unsigned magnetic flux, $\phi$, and thus a much larger $|\bar\hv|$.

Noteworthy, $\hjprime$ in our study appears indicative only for the upcoming major eruptive flares in AR~11158, but not for the major confined flares in AR~12192. This indicates that $|\hj|/|\hv|$ is a good proxy for the eruptive potential of an AR, but cannot be expected to serve as an indicator whether an upcoming flare will involve the rearrangement of the magnetic field in a confined (non-eruptive) or eruptive manner. This speculation is supported by the recent work of \cite{2019arXiv190706365M} who studied the magnetic helicity of AR~12673, around two consecutive major X-class flares (a preceding confined and a following eruptive one, about three hours later). They found values of $|\hj|/|\hv|$ comparable with that of AR~11158 in our study, with even higher values prior to the major confined flare.  

Our observation based analysis represents an extension of the work by \cite{2018ApJ...855L..16J} and \cite{2019arXiv190706365M}, who suggested values of $|\hj|/|\hv|\gtrsim0.15$ to be characteristic for the immediate pre-flare magnetic field, based on NLFF modeling of the solar corona above selected ARs. Based on our long-term analysis of AR~11158, we find values of $|\bar\hj|/|\bar\hv|\gtrsim0.17$ prior to the major eruptive flares. In addition, we notice that the pre-flare magnitude of $|\hj|/|\hv|$ appears unrelated to the intensity of the eruptive flares.

In the statistical survey of the magnetic helicity injection in (345) 48 (non-) X-class flare productive ARs by \cite{2007ApJ...671..955L}, $\hv$ was approximated by the accumulated photospheric helicity flux during specified observing intervals. From their Fig.~8, a significant spread of $\hv$ is noticeable for a given AR magnetic flux, and that the corresponding AR may not necessarily produce an X-flare. Since literally all X-class flares are eruptive \citep[\eg,][]{2006ApJ...650L.143Y}, this finding is equivalent to the argument that the normalized helicity, $\hv/\phi^2$, is not indicative for eruptivity. In our work, we find values for $\hvprime$ for non-eruptive AR 12192 in the same range as those of eruptive AR~12673 \citep{2019arXiv190706365M}. Thus, in line with the statistical work of \cite{2007ApJ...671..955L}, we suggest that $\hvprime$ does not serve as discriminant factor for the eruptive potential of a solar AR.

For completeness, we note distinct local maxima in the time profile of $\efprime$ prior to eruptive flare occurrences in AR~11158, though small compared to the corresponding variations for AR~12192, and with much less pronounced differences than for the respective time profiles of $\hjprime$. Therefore, we agree with earlier works \citep[\eg,][]{2017A&A...601A.125P,2019arXiv190706365M} that though $\Efree$ (and thus $\Efree/\Etot$) is tightly linked to the potential eruptivity of an AR, it does not represent a sufficient condition for an eruption to occur.

Last, we note that all of the analyzed extensive quantities (and possibly also the intensive ones) may depend, in general, on the extension of the analyzed volume and the spatial resolution of the vector magnetogram data. In the present work, for convenience, we binned the photospheric vector magnetic field data by a factor of four, prior to magnetic field modeling and subsequent magnetic helicity computation. A first attempt to quantify corresponding differences has been presented by \cite{2015ApJ...811..107D}, who applied different existing NLFF modeling techniques to a sequence of vector magnetograms with different spatial resolutions, constructed from polarimetric inversion of polarization spectra that were binned by factors ranging from 2 to 16. Their results suggested that, even given a sufficient fulfillment of the solenoidal property, the magnetic helicity computed from the model magnetic fields of different spatial resolution \citep[using the method of][]{2012SoPh..278..347V}, for a given NLFF method, may vary substantially.

We expect further substantiation and clarification of the aspects discussed above from anticipated future studies, based on the analysis of the helicity budgets of a large number of solar ARs, that will correspondingly allow more robust statements.

\acknowledgements
{\footnotesize
We thank the anonymous referee for careful consideration of this manuscript and helpful comments. J.\,K.\,T.\ was supported by the Austrian Science Fund (FWF): P31413-N27. E.\,P., K.\,M., and L.\,L.\ acknowledge support of the French Agence Nationale pour la Recherche through the HELISOL project ANR-15-CE31-0001. G.\,V.\ acknowledges support of the Leverhulme Trust Research Project Grant 2014-051. K.\,D.\ gratefully acknowledges the support of the french Centre National d'\'Etudes Spatiales. {\it SDO} data are courtesy of the NASA/{\it SDO} AIA and HMI science teams. This article profited from discussions during the meetings of the ISSI International Team {\it Magnetic Helicity in Astrophysical Plasmas}.
}


\end{document}